\documentclass[sigconf]{acmart}

\AtBeginDocument{%
  }

\copyrightyear{2025}
\acmYear{2025}
\setcopyright{rightsretained}
\acmConference[WWW Companion '25]{Companion Proceedings of the ACM Web Conference 2025}{April 28-May 2, 2025}{Sydney, NSW, Australia}
\acmBooktitle{Companion Proceedings of the ACM Web Conference 2025 (WWW Companion '25), April 28-May 2, 2025, Sydney, NSW, Australia}
\acmDOI{10.1145/3701716.3717559}
\acmISBN{979-8-4007-1331-6/2025/04}

\begin{document}

\title{The 1$^{st}$ EReL@MIR Workshop on Efficient Representation Learning for Multimodal Information Retrieval}

\author{Junchen Fu}
\authornote{Equal contribution.}
\affiliation{%
  \institution{University of Glasgow}
  \city{Glasgow}
  \country{UK}
}
\email{j.fu.3@research.gla.ac.uk}

\author{Xuri Ge}
\authornotemark[1]

\affiliation{%
  \institution{Shandong University}
  \city{Shandong}
  \country{China}
  }
\email{xurigexmu@gmail.com}
\author{Xin Xin}
\affiliation{%
  \institution{Shandong University}
  \city{Shandong}
  \country{China}}
\email{xinxin@sdu.edu.cn}

\author{Haitao Yu}
\affiliation{%
  \institution{University of Tsukuba}
  \city{Tsukuba}
  \country{Japan}}
\email{yuhaitao@slis.tsukuba.ac.jp}

\author{Yue Feng}
\affiliation{%
  \institution{University of Birmingham}
  \city{Birmingham}
  \country{UK}}
\email{y.feng.6@bham.ac.uk}

\author{Alexandros Karatzoglou}
\affiliation{
\institution{Amazon}\streetaddress{}\city{Barcelona}\country{Spain}}
\email{alexandros.karatzoglou@gmail.com}

\author{Ioannis Arapakis}
\affiliation{
\institution{Telef\'{o}nica Scientific Research}\streetaddress{}\city{Barcelona}\country{Spain}}
\email{arapakis.ioannis@gmail.com}

\author{Joemon M. Jose}
\affiliation{%
  \institution{University of Glasgow}
  \city{Glasgow}
  \country{United Kingdom}}
\email{joemon.jose@glasgow.ac.uk}

\renewcommand{\shortauthors}{Junchen et al.}

\begin{abstract}
  Multimodal representation learning has garnered significant attention in the AI community, largely due to the success of large pre-trained multimodal foundation models like LLaMA, GPT, Mistral, and CLIP. These models have achieved remarkable performance across various tasks of multimodal information retrieval (MIR), including web search, cross-modal retrieval, and recommender systems, etc. However, due to their enormous parameter sizes, significant efficiency challenges emerge across training, deployment, and inference stages when adapting these models' representation for IR tasks. 
  These challenges present substantial obstacles to the practical adaptation of foundation models for representation learning in information retrieval tasks.
  
To address these pressing issues, we propose organizing the first EReL@MIR workshop at the Web Conference 2025, inviting participants to explore novel solutions, emerging problems, challenges, efficiency evaluation metrics and benchmarks. This workshop aims to provide a platform for both academic and industry researchers to engage in discussions, share insights, and foster collaboration toward achieving efficient and effective representation learning for multimodal information retrieval in the era of large foundation models. 
\end{abstract}




\maketitle

\section{BACKGROUND AND OBJECTIVES}
Multimodal representation learning plays an important role in real-world multimedia applications, e.g., e-commerce platforms with multimodal recommendation systems (Amazon, eBay, Alibaba, etc.), or search engines with multimodal retrieval strategies (Google, Bing, Baidu, etc.).
In recent years, it has attracted increasing research attention within the AI community, primarily due to the success of large pre-trained multimodal foundation models (MFMs) like LLaMA \cite{touvron2023llama}, GPT \cite{achiam2023gpt}, Mistral \cite{jiang2023mistral}, CLIP \cite{radford2021learning}, etc.  
These MFMs have set new benchmarks in various domains for multimodal information retrieval (MIR), such as web search \cite{wei2023uniir}, cross-modal retrieval \cite{wang2024unified,long2024cfir}, and recommender systems \cite{geng2023vip5,yuan2023go,liu2024multimodal}.

Despite the impressive performance of foundation models, adapting them to real-world MIR applications poses unique challenges, particularly in representing diverse modalities and adapting large pre-trained MFMs for efficient retrieval and recommendation tasks \cite{geng2023vip5,liu2024rec,yuan2024asking,fu2024exploring,fu2024iisan,yuan2023go,fu2024efficient}. The existing MIR research typically requires substantial computational resources, leading to higher VRAM (video random access memory) usage and increased costs for both training and maintenance. Additionally, real-world MIR tasks impose strict latency requirements, demanding highly efficient inference. This necessitates models that can effectively handle multiple modalities and deliver fast, responsive results within tight computational limits.

The current state of research has primarily focused on improving the accuracy using these multimodal foundation models, but less attention has been given to the efficiency aspects \cite{hou2024large,wu2024survey,zhu2023large,liu2024multimodal,liao2024llara}.
This underscores the urgent need for solutions that can balance both performance and efficiency in adapting these models for information retrieval, specifically multimodal IR tasks. Additionally, the lack of standardised metrics and benchmarks for evaluating the efficiency of the MIR approaches results in more complicated and inconsistent development of the research community.

To address these pressing issues, the EReL@MIR workshop at the Web Conference 2025 will provide a dedicated platform for researchers and practitioners to explore novel solutions, discuss emerging problems, and propose new benchmarks and evaluation metrics. The Web Conference is one of the leading conferences with numerous recognized research works focusing on IR-related research and applications. Our EReL@MIR workshop is highly relevant to the conference. We believe that organizing the EReL@MIR workshop with the Web Conference 2025 can achieve the following three objectives: (i) stimulate interesting research of efficient multimodal learning to meet complex IR demands; (ii) encourage collaboration on unified new metrics and benchmarks for assessing efficiency; and (iii) drive progress toward practical solutions that make multimodal IR more feasible for real-world application.

\vspace{-0.1in}

\section{TOPICS AND THEMES}
As the first EReL@MIR workshop, we will focus on the \textbf{fundamental challenges} in multimodal representation learning for MIR, addressing the efficiency issues in the practical deployment of large MFMs.

\noindent \textbf{1. Efficient Multimodal Representation Adaptation based on MFMs.}  
Adapting large multimodal foundation models, such as LLaMA, GPT, and CLIP, for specific IR tasks requires significant computational resources, especially when learning complex representations across modalities. The challenge is to reduce the computational cost of adapting these models while retaining high-quality multimodal representations. Approaches such as parameter-efficient fine-tuning or model compression are crucial for ensuring the scalability of these models in real-world IR systems.

\noindent \textbf{2. Data-Efficiency in Multimodal Representation Learning.}  
Training multimodal foundation models on large, diverse datasets is highly resource-intensive, requiring the learning of complex representations across multiple modalities, such as text, images, and videos. To address this, efficient data utilization techniques—like strategic data sampling, knowledge distillation, and data denoising—can significantly reduce the data volume needed to develop effective multimodal representations. Tackling this challenge is crucial to lowering training costs and accelerating model development, while maintaining high-quality representations.

\noindent \textbf{3. Efficient Multimodal Fusion for Representation Learning.}  
The integration of different modalities (e.g., text, image, video, and audio) into a unified representation remains computationally expensive and is a major challenge in multimodal learning. Developing more efficient fusion mechanisms—such as lightweight architectures or adaptive fusion techniques—can reduce the computational burden while ensuring high-quality multimodal representations. This challenge is critical to making multimodal fusion feasible in practical IR scenarios.

\noindent \textbf{4. Efficient Cross-Modality Interaction for MIR.}
The effective interactions among modalities like text, images, and videos is a key to strong multimodal representations. However, cross-modality interactions often rely on computationally intensive techniques like deep attention mechanisms, GNNs, or contrastive learning, all of which increase training costs. The challenge is to develop lightweight interaction mechanisms that reduce these computational burdens while still ensuring effective information exchange between modalities. We call for approaches that address this issue by achieving efficient cross-modality interaction without the heavy overhead typically associated with current methods.

\noindent \textbf{5. Real-Time Inference for Multimodal Representations.}  
Real-time IR applications, including search engines and recommender systems, require rapid inference of multimodal representations to ensure a responsive user experience. However, the inference speed of large models can be a bottleneck due to the complexity of the learned representations. Optimizing real-time inference through techniques such as model pruning, quantization, and efficient encoding methods like vector quantization is essential for deploying these models in latency-critical IR applications.

\noindent \textbf{6. Efficient MIR Foundation Models}. Current foundation models for representation learning predominantly rely on transformer-based architectures. While these models demonstrate strong performance, they are hindered by their enormous parameter sizes and significant computational overhead. Therefore, there is a pressing need for more efficient solutions tailored to MIR tasks, which can mitigate these challenges without sacrificing effectiveness.

\noindent \textbf{7. Benchmarks and Metrics for Multimodal Representation Learning Efficiency.}  
Currently, no standardized benchmark or widely recognized comprehensive metric exists to specifically evaluate the efficiency of representation learning in MIR tasks. Efficiency involves multiple factors, such as VRAM usage, training and inference time, and the number of parameters, etc. The absence of these resources makes it difficult to assess and compare different models in terms of their efficiency. Establishing benchmarks and metrics that evaluate both the effectiveness and efficiency of multimodal representation learning is essential for guiding future research and fostering innovation in the field.

\section{FORMAT}



\subsection{Target Audience and Accpeted Submissions}
This workshop’s appeal lies in its strong focus on the evolving field of information retrieval and recommender systems. It is expected to attract a diverse audience, including researchers, industry professionals, and academics. Through our rigorous review process, we ultimately accepted 9 high-quality papers, with 8 of them included in the proceedings.

Papers submitted to the workshop underwent a standard peer-review process. Each submission was evaluated by at least three reviewers, and accepted papers were selected based on originality, relevance, and technical soundness. The final decision was made by program committee members in collaboration with the organizers.

\subsection{Workshop Advertisement}
To ensure wide participation, the workshop will be promoted through various channels, including social media platforms (X, LinkedIn, Wechat, REDnote) and academic mailing lists. Organizers will also reach out directly to researchers working in the field of multimodal learning and information retrieval.

\section{Related Workshop}
\textbf{List of related workshops:}
\begin{itemize}
    \item Multimodal KDD 2023: International Workshop on Multimodal Learning (KDD 2023)\footnote{\url{https://multimodal-kdd-2023.github.io/}}
    \item Multimodal Learning and Applications Workshop (ECCV2018, CVPR 2019-2024)\footnote{https://mula-workshop.github.io/}
    \item Workshop on Multimodal Search and Recommendations (CIKM 2024)\footnote{\url{https://cikm-mmsr.github.io/}}
    \item Workshop on Multimodal Representation Learning (MRL): Perks and Pitfalls (ICLR 2023)\footnote{\url{https://mrl-workshop.github.io/iclr-2023/}}
    \item Workshop on Representation Learning for NLP (ACL 2016-2024)\footnote{\url{https://aclanthology.org/events/repl4nlp-2024/}}
\end{itemize}

The EReL@MIR Workshop on Efficient Representation Learning for Multimodal Information Retrieval at The Web Conference 2025 focuses on the efficient representation learning of multimodal information retrieval tasks. While workshops such as Multimodal KDD 2023, Multimodal Learning and Applications Workshop,  Multimodal Search and Recommendations, and MRL Workshop cover a broad range of multimodal applications, EReL@MIR specifically focuses on the challenge of efficiency in representation learning for MIR. Unlike these workshops that prioritize performance, EReL@MIR centers on the less explored but crucial aspect of computational efficiency, which is essential for real-world application and deployment. The workshop will bring together researchers to discuss solutions that balance efficiency and performance, ensuring scalable applications of multimodal models in IR tasks.
\section{List of Programme Committee Members}
\begin{itemize}
    \item Zhiwei Chen - Jiangxi Normal University, China
    \item Mingyue Cheng - University of Science and Technology of China, China
    \item Wenhao Deng - Westlake University, China
    \item Shichang Feng - Nanjing University of Aeronautics and Astronautics
    \item Keji He - Shandong University, China
    \item Hengchang Hu - National University of Singapore, Singapore
    \item Jiayi Ji - National University of Singapore, Singapore
    \item Hui Li - Xiamen University, China
    \item Linqing Li - Central China Normal University, China
    \item Qian Li - Beijing University of Posts and Telecommunications, China
    \item Siwei Liu - University of Aberdeen, UK
    \item Xi Wang - University of Sheffield, UK
    \item Songpei Xu - University of Glasgow, UK
    \item Zheng Yuan - The Hong Kong Polytechnic University, Hong Kong
    \item Jiaqi Zhang - University of Queensland, Australia
    \item Kaiwen Zheng - University of Glasgow, UK
    \item Zhiwei Zheng - University of Glasgow, UK
\end{itemize}

\section{Diversity Aspects}
This workshop emphasizes diversity across \textbf{gender, institutional affiliation, and nationality} in its \textbf{speakers, organizers, and Programme Committee}. Therefore, the invited speakers will represent a wide range of institutions and nationalities, ensuring diverse perspectives. The organizers also reflects a commitment to diversity, with both male and female members, and institutional representation from eight organizers affiliated with six different institutions across the UK, Spain, China, and Japan, covering both academia and industry. In terms of nationality, the team includes individuals from the UK, the European Union, and China, ensuring a maximum dirveristy. The Programme Committee members are also selected to ensure high diversity across the aforementioned three aspects.
\vspace{-0.1in}

\section{Explaination for Relevance:} Multimodal representation learning, driven by large foundation models, has reshaped the landscape of MIR, enabling integration across modalities such as text, images, and video. However, the efficiency challenges—such as increased computational demands—limit their practical application in IR systems. The Web Conference, renowned for advancing web technologies and IR, serves as an ideal venue for addressing these challenges. The EReL@MIR workshop is highly relevant to The Web Conference, aligning closely with its focus by advancing efficient representation learning methods for MIR tasks.
\vspace{-0.1in}

\section{OGANIZERS INFORMATION}
\noindent \textbf{Prof. Joemon M. Jose} is a Professor at the School of Computing Science, University of Glasgow, Scotland and a member of the Information Retrieval group. His research focuses around the following three themes: (i) Social Media Analytics; (ii) Multi-modal interaction for information retrieval; (iii) Multimedia mining and search. He has published over 300 papers with more than 10,000 Google Scholar citations, and an h-index of 51. He leads the GAIR Lab which investigates research issues related to the above themes. He has been serving as the program committee chair and member for numerous top international conferences (e.g., SIGIR, WWW, and ECIR). He also serves as a PC chair for SIGIR-AP 2024.

\noindent \textbf{Dr. Alexanderos Karatzgolou} is a Principal Applied Scientist at Amazon. His contributions, such as GRU4Rec and kernlab, have significantly influenced and advanced future research in machine learning and recommender systems. His work has garnered over 17,000 Google Scholar citations, with an h-index of 42. Before joining Amazon, he was a Staff Research Scientist at Google DeepMind. He also served as Director of Telefonica Research in Barcelona and holds a PhD from the Vienna University of Technology.

\noindent \textbf{Dr. Ioannis Arapakis} is a Principal Research Scientist at Telefónica Research, specializing in behavior interpretation algorithms for user modeling in both offline and online contexts, with a focus on web search. He holds a Ph.D. in Information Retrieval from the University of Glasgow. He has published papers in leading international conferences and journals, such as SIGIR, WSDM, CIKM, and TOIS. Previously, he was a Research Scientist at Yahoo Labs, where he worked on data mining, IR, HCI, and multimedia mining projects.

\noindent \textbf{Dr. Xin Xin} is a Tenure-Track Assistant Professor at the School of Computer Science and Technology of Shandong University. Before that, he earned his Ph.D. degree from the University of Glasgow. His current research interests include information retrieval, natural language processing, and reinforcement learning. He has published more than 40 papers in top conferences (e.g., WWW, SIGIR, ACL, WSDM) and journals (e.g., TOIS, TKDE), and received the Best Paper Honor Mention at WSDM 2024. He has organized the DRL4IR workshop in SIGIR, KEIR workshop in ECIR, and R$^{3}$AG workshop in SIGIR-AP.

\noindent \textbf{Dr. Haitao Yu} is a Tenured Associate Professor at University of Tsukuba and leading the Information Intelligence research group. His research focuses on Information Retrieval, Knowledge Graph, and Machine Learning. He has published numerous papers on top international conferences (e.g., WSDM, CIKM, SIGIR, WWW, ECIR, and AAAI) and journals (e.g., Information Processing and Management, and Information Retrieval Journal). He is the co organizer of the NTCIR tasks of Temporalia-2 and AKG. He has been serving as the program committee member for numerous top international conferences (e.g., WSDM, CIKM, SIGIR, and ECIR).

\noindent \textbf{Dr. Yue Feng} is an assistant professor at the University of Birmingham. She got her Ph.D. from University College London. Her research interests lie in natural language processing and information retrieval. She won the Alexa Prize TaskBot Challenge in 2021 and was awarded the Baidu Excellent Research Intern Star in 2020. She served as an Area Chair at the NLPCC’24 conference.

\noindent \textbf{Dr. Xuri Ge} is now a tenure-track assistant professor in the school of artificial intelligence, Shandong University. He earned his PhD at the University of Glasgow (UK) and received M.S. degree from Xiamen University (China). His current research interests include computer vision, multimodal representation learning, and information retrieval. He has contributed to several leading conferences and journals, including NeurIPS, SIGIR, ACM MM, CIKM, ACM TIST, and IP\&M, ect. He also serves as the PC member and reviewer for tier-1 conferences and journals, such as NeurIPS, WWW, MM, AAAI, ICLR, TKDE, TOIS, IJCV, etc. He has organized the 3DMM Workshop in ICME and R$^{3}$AG workshop in SIGIR-AP.

\noindent \textbf{Junchen Fu} is a PhD student under the supervision of Prof. Joemon Jose at the School of Computing Science, University of Glasgow. Prior to this, he worked as a Research Assistant at Westlake University, where he specialized in modality-based recommender systems. He holds an MSc in Computer Science from The Chinese University of Hong Kong, graduating with Dean's List honors. He has contributed to several leading AI conferences and journals, including SIGIR, WWW, WSDM, MM, and TPAMI. He also serves as the PC member and reviewer for top tier conferences and journals, such as ICLR, SIGIR, SIGKDD, WWW, WSDM, MM,  TKDE, TMM, TOIS, etc. His research primarily focuses on enhancing the efficiency of adapting multimodal foundation models for recommender systems.

\bibliographystyle{ACM-Reference-Format}
\bibliography{sample-base}

\end{document}